\documentclass[galaxies,article,accept,moreauthors,pdftex]{Definitions/mdpi}

\firstpage{1} 
\pubvolume{9}
\issuenum{3}
\articlenumber{68}
\pubyear{2021}
\copyrightyear{2021}
 \externaleditor{Academic Editor: Elena Moretti and Francesco Longo} % For journal Automation, please change Academic Editor to "Communicated by" %MDPI please add this part. 
\datereceived{30 July 2021} 
\dateaccepted{13 September 2021} 
\datepublished{15 September 2021} 
\hreflink{https://doi.org/10.3390/\\galaxies9030068} % If needed use \linebreak
\Title{The Low-Energy Spectral Index of Gamma-Ray Burst Prompt Emission from Internal Shocks}

\TitleCitation{The Low-Energy Spectral Index of Gamma-Ray Burst Prompt Emission from Internal Shocks}

\Author{Kai Wang $^{1,}$*\orcidA{} and  Zi-Gao Dai $^{2,3}$} %MDPI \hl{} %Please carefully check the accuracy of names and affiliations. 

\AuthorNames{Kai Wang and Zi-Gao Dai}

\AuthorCitation{Wang, K.; Dai, Z.-G.}
\address{%
$^{1}$ \quad Department of Astronomy, School of Physics, Huazhong University of Science and Technology, Wuhan~430074, China \\
$^{2}$ \quad Department of Astronomy, University of Science and Technology of China, Hefei 230026, China; daizg@ustc.edu.cn\\
$^{3}$ \quad School of Astronomy and Space Science, Nanjing University, Nanjing 210023, China}

\corres{Correspondence: kaiwang@hust.edu.cn}

\abstract{The prompt emission of most gamma-ray bursts (GRBs) typically exhibits a non-thermal Band component. The synchrotron radiation in the popular internal shock model is generally put forward to explain such a non-thermal component. However, the low-energy photon index $\alpha \sim -1.5$ predicted by the synchrotron radiation is inconsistent with the observed value $\alpha \sim -1$. Here, we investigate the evolution of a magnetic field during propagation of internal shocks within an ultrarelativistic outflow, and revisit the fast cooling of shock-accelerated electrons via synchrotron radiation for this evolutional magnetic field. We find that the magnetic field is first nearly constant and then decays as $B'\propto t^{-1}$, which leads to a reasonable range of the low-energy photon index, $-3/2 < \alpha < -2/3$. In addition, if a rising electron injection rate during a GRB is introduced, we find that $\alpha$ reaches $-2/3$ more easily. We thus fit the prompt emission spectra of GRB 080916c and GRB~080825c.}


\begin{document}
%%%%%%%%%%%%%%%%%%%%%%%%%%%%%%%%%%%%%%%%%%
\section{Introduction}           %% first-level sections will be auto-capitalized
\label{sect:intro}

The prompt radiation mechanism of gamma-ray bursts (GRBs) is still being debated, even though the prompt spectra can usually be fitted well by the Band function~\citep{ban93}, which suggests a smoothly jointed broken power law with the low-energy photon index $\alpha \sim -1$, the high energy photon index $\beta \sim -2.2$ and the peak energy $E_{p} \sim 250\,\mathrm{keV}$~\citep{pre00,gru14}. Currently, neither the possible one-temperature thermal emission from an ultrarelativistic fireball, nor the single synchrotron radiation from shock-accelerated electrons within this fireball, provide an explanation for such a low-energy photon index (see~\citep{zhang14,kumar14} for a review).

Generally, there are two mechanisms that explain the low energy photon index $\alpha \sim -1$ of the GRB prompt emission. The first mechanism is the Comptonized quasi-thermal emission from the photosphere of an ultrarelativistic outflow~\citep{tho94,mes00,mes02,rees05,peer06a,tho07,peer08,gia08,bel10,laz10,peer11,lun13,laz13,deng14}. The second mechanism is synchrotron and/or synchrotron self-Compton (SSC) emission in the optically thin region. For fast-cooling synchrotron radiation in the internal shock model, possible solutions include invoking a small-scale rapidly decaying magnetic field~\citep{peer06b}, a decaying magnetic field with a power-law index in a relativistically-expanding outflow~\citep{uhm14,zhang16,geng18}, a decaying magnetic field in a post-shock region~\citep{zhao14}, Klein--Nishina (KN) cooling~\citep{wang09,daigne11}, an adjustable synchrotron self-absorption frequency~\citep{lloyd00,shen09}, or the acceleration process~\cite{xu18} and other evolutional model parameters~\citep{bos14}. Alternatively, slow cooling was introduced to understand the low-energy photon index~\citep{burgess14}. In addition to the internal shock model, the other energy dissipation mechanisms, such as the ICMART model~\citep{zhang11a}, were proposed to solve the low-energy spectral index issue. In some models (e.g.,~\citep{pan00}), the observed prompt emission of GRBs is understood to be dominated by the SSC emission, while the synchrotron radiation is in much lower energy bands.

The fast cooling synchrotron radiation in the internal shock model is generally considered to be a straightforward and leading mechanism to explain the GRB prompt emission spectra, and the most important issue in this model is to explain the low-energy spectral index. An underlying assumption in the traditional synchrotron internal shock model is to calculate the electron cooling without considering the evolution of the magnetic field. In other words, the magnetic field is treated as a constant and its effect in the continuity equation of electrons is usually ignored (e.g.,~\cite{sar98}). In the fast cooling case, the predicted low-energy photon index $\alpha \sim -3/2$ is much softer than observed. In this paper, we try to alleviate this problem. We calculate the magnetic field in the realistic internal shock model during a collision of two relativistic thick shells and obtain an evolutional form of the magnetic field, $B'\propto$ constant before the time $\delta t$ that is nearly equal to the ejection time interval of the two shells, and $B'\propto t^{-1}$ after the time $\delta t$. We consider the cooling of electrons accelerated by internal shocks for this evolutional magnetic field, and find the resulting spectral index $\alpha \sim -3/2$ for $B' \propto $ constant and $\alpha \sim -2/3$ for $B' \propto t^{-1}$, by adopting a cooling method similar to that in Ref.~\cite{uhm14}. Actually, these two cases may coexist, and the outflow may undergo the first case and then the second case, so theoretically the actual index $\alpha$ will range from $-3/2$ to $-2/3$. Furthermore, below the peak energy $E_p$ there is a gradual process, so that $\alpha$ is only close to $-1$. In addition, we consider a rising electron injection rate, leading to a larger $\alpha$, slightly smaller than $-2/3$.

This paper is organized as follows. We calculate the dynamics of a collision between two thick shells in Section \ref{sec2}. In Section \ref{sec3}, we investigate the electron cooling and its synchrotron radiation with an evolutional magnetic field and a rising electron injection rate. In the final Section, discussions and conclusions are given.

\section{Dynamics of Two-Shell Collision}\label{sec2}
In the popular internal shock model, an ultrarelativistic fireball consisting of a series of
shells with different Lorentz factors can produce prompt emission
through collisions among these shells. For the dynamics of two-shell collision, we adopt the same approach as the one in~\cite{wang13}. In order to present one GRB prompt emission component (with duration $\sim$ few seconds), we here consider two thick shell--shell collision to produce a consistent GRB pulse with a duration of few seconds (i.e., the slow pulse) and the fast pulses with a duration of $\sim 0.01 \rm s$ in GRBs may be caused by the 
density fluctuation of the shell. Under this assumption, a prior slow thick shell A with bulk lorentz factor
$\gamma_{A}$ and kinetic luminosity $L_{k,A}$, and a
posterior fast thick shell B with bulk lorentz factor $\gamma_{B}$
(where $\gamma_{B}>\gamma_{A}\gg 1$) and kinetic luminosity $L_{k,B}$ is adopted.
The collision of the two shells begins at radius~\cite{wang13} 
\begin{align}{}
{R_{col}} & = {\beta _B}c\frac{{{\beta _A}\Delta {t_{int}}}}{{({\beta _B} -
{\beta _A})}} \simeq \frac{{2\gamma _A^2c\Delta {t_{int}}}}{{1 -
{{({\gamma _A}/{\gamma _B})}^2}}} \nonumber \\
& \equiv  2\gamma _A^2c\delta t \simeq 5.4 \times {10^{14} \gamma _{A,2.5}^2 \delta t_{,-1}} \; \mathrm{cm},
\end{align}
where $\Delta {t_{int}}$ is the time interval between the two thick shells,
and $\delta t\equiv \Delta {t_{int}}/[1 -(\gamma_A/\gamma_B)^2]$ is a redefined time interval. For $\gamma_{A}\ll\gamma_{B}$, $\delta t\simeq\Delta {t_{int}}$. The conventional expression $Q_{,m}=Q/10^m$ is used. During the collision, there
are four regions separated by internal forward-reverse shocks: (1) the
unshocked shell A; (2) the shocked shell A; (3) the shocked shell B;
and (4) the unshocked shell B, where regions 2 and 3 are separated by
a contact discontinuity. %MDPI is the bold necessary? please confirm. the same below. PLease check all bold in text, if not necessary, please remove it.

The particle number density of a shell measured in its comoving
frame can be calculated as~\cite{yu09}:
\begin{equation}
{n'_i} = \frac{{{L_{k,i}}}}{{4\pi {R^2}\gamma _i^2{m_p}{c^3}}},
\end{equation}
where $R$ is the radius of the shell and subscript $i$ can be taken
as A or B. As in the literature~\citep{sari95,kumar00,dai02,yu09,wang13}, we derive the dynamics of
internal forward-reverse shocks. In order to get a high prompt emission
luminosity, it is reasonable to assume $\gamma_{A}\ll\gamma_{B}$ and
$L_{k,A}=L_{k,B}\equiv L_{k}$. Assuming that $\gamma_{1}$,
$\gamma_{2}$, $\gamma_{3}$, and $\gamma_{4}$ are Lorentz factors of
regions 1, 2, 3 and 4 respectively, we have
$\gamma_{1}=\gamma_{A}$, $\gamma_{4}=\gamma_{B}$, and $n'_1\gg
n'_4$. If a fast shell with low particle
number density catches up with a slow shell with high particle
number density and then they collide with each other, a Newtonian
forward shock (NFS) and a relativistic reverse shock (RRS) may be
generated~\citep{yu09,wang13}. So we can obtain
$\gamma_{1}\simeq\gamma_{2}=\gamma_{3}=\gamma\ll\gamma_{4}$. Then,
according to the jump conditions between the two sides of a shock
\citep{bla76}, the comoving internal energy densities of the two
shocked regions can be calculated following ${e'_2} = ({\gamma _{21}} -
1)(4{\gamma _{21}} + 3){n'_1}{m_p}{c^2}$ and ${e'_3} = ({\gamma _{34}} -
1)(4{\gamma _{34}} + 3){n'_4}{m_p}{c^2}$, where
$\gamma_{21}=\frac{1}{2}({\gamma _1}/{\gamma _2} + {\gamma
_2}/{\gamma _1})$ and $\gamma_{34}=\frac{1}{2}({\gamma _3}/{\gamma
_4} + {\gamma _4}/{\gamma _3})$ are the Lorentz factors of region 2
relative to the unshocked region 1, and region 3 relative to region
4, respectively. It is required that $e'_2=e'_3$ because of the
mechanical equilibrium. We have~\cite{yu09,wang13}
\begin{equation}
\frac{{({\gamma _{21}} - 1)(4{\gamma _{21}} + 3)}}{{({\gamma _{34}} - 1)
(4{\gamma _{34}} + 3)}} = \frac{{n'_4}}{{n'_1}}
={\left( {\frac{{{\gamma _1}}}{{{\gamma _4}}}} \right)^2} \equiv f.
\end{equation} %MDPI please confirm whether need indent.
Two relative Lorentz factors can be calculated as ${\gamma _{21}}
\approx \frac{{f\gamma _4^2}}{{7\gamma _1^2}} + 1 = \frac{8}{7}$,
and ${\gamma _{34}} = \frac{{{\gamma _4}}}{{2{\gamma _1}}} \gg 1$.
Assuming that $t$ is the observed shell--shell interaction time since the
prompt flare onset, the radius of the system during the collision can
be written as
\begin{equation}
R = {R_{col}} + 2{\gamma ^2}ct \simeq 2{\gamma ^2_1}c(t + \delta t).
\label{R}
\end{equation}

During the propagation of the shocks and before the shock crossing time,
the instantaneous electron injection numbers (in $dt$) in regions 2 and 3 can be calculated as follows~\citep{dai02}:
\begin{equation}
{dN_{e,2}} = 8\pi {R^2}{n'_1}({\gamma _{21}}{\beta _{21}}/\gamma
\beta ){\gamma ^2}c dt
\label{dNe2}
\end{equation}
and
\begin{equation}
{dN_{e,3}} = 8\pi {R^2}{n'_4}({\gamma
_{34}}{\beta _{34}}/\gamma \beta ){\gamma ^2}c dt ,
\label{dNe3}
\end{equation}
respectively.
%For a specific collision, $dN_{e,2}$ and $dN_{e,3}$ are constants per unit time.

\section{Synchrotron Radiation with a Decaying Magnetic Field and a Variable Electron Injection Rate}\label{sec3}

\subsection{Synchrotron Radiation with a Decaying Magnetic Field}

\begin{figure}[H]
 \includegraphics[width=10.5 cm]{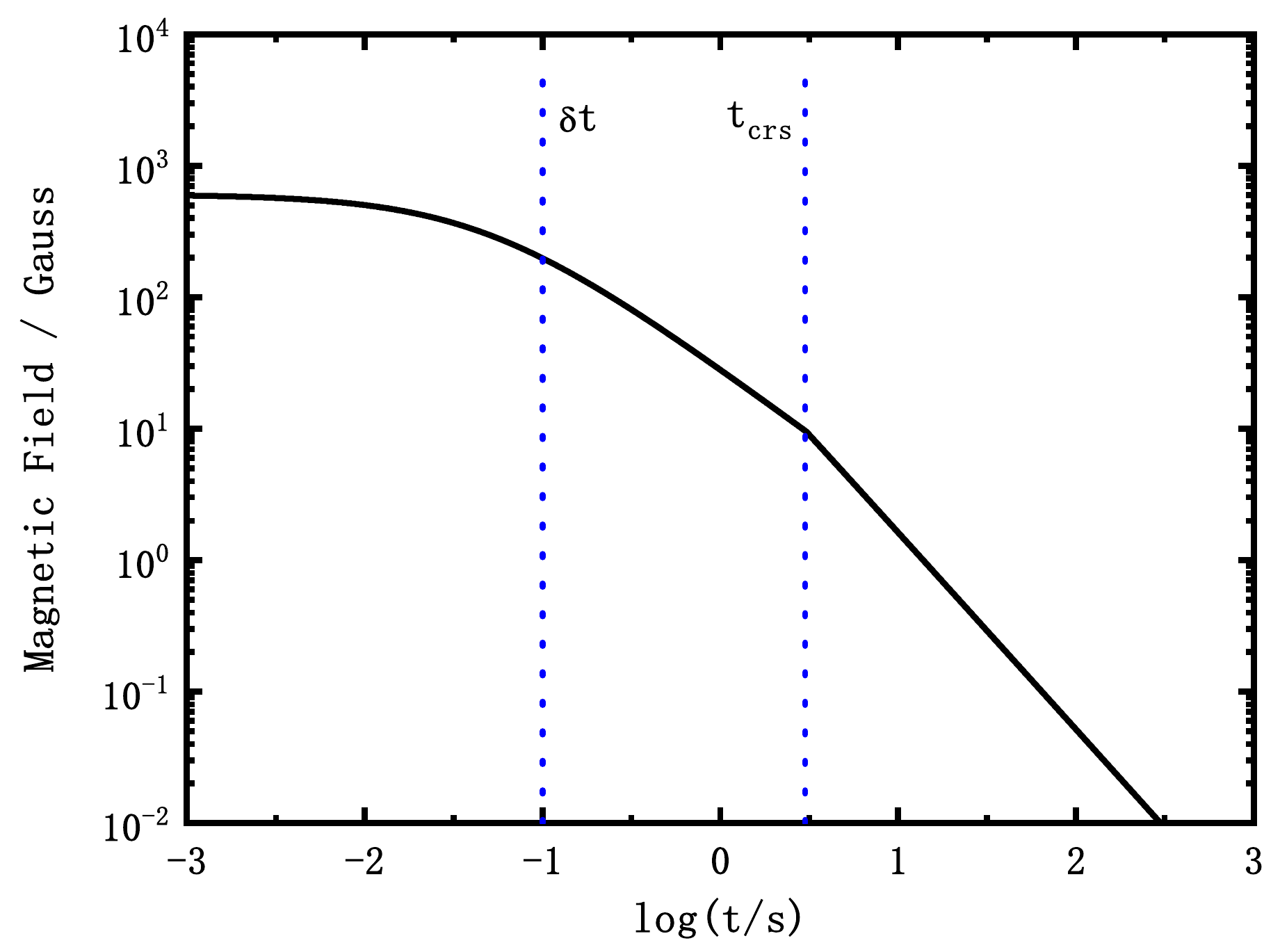}
  \caption{The magnetic field as a function of time. The two blue vertical dotted line represents the redefined interval $\delta t =0.1 \; \mathrm{s}$ and the shock crossing time $t_{crs}=3 \; \mathrm{s}$, respectively. After the shock crossing time, the merged shell expands adiabatically and $s = 3$ is assumed. The dynamics parameters $L_{k}=10^{51} \; \mathrm{erg \, s^{-1}}$, $\gamma_{1}=300$, $\gamma_{4}=30000$, $p=2.5$, $\epsilon_e=0.3$, $\epsilon_B=0.3$, and $z=1$ are taken from numerical calculations.\label{f1}}
\end{figure}

As usual, we assume that fractions $\epsilon_{B}$ and
$\epsilon_{e}$ of the internal energy density in a GRB shock are
converted into the energy densities of the magnetic field and
electrons, respectively. Thus, using ${B'_i} = {(8\pi {\epsilon
_B}{e'_i})^{1/2}}$ for $i=$2 or 3, we can calculate the strength of the magnetic field before the shock crossing time $t_{crs}$ by
\begin{equation}
{B'_2} ={B'_3} = {\left[\frac{{{\epsilon_B}L_k}}{{2{\gamma
^6_1}{c^3}{{(t + \delta t)}^2}}}\right]^{1/2} },
\label{b}
\end{equation}
and find that the change of the magnetic field before $\delta t$ can be ignored (i.e., $B'_i \propto {\rm constant}$), but after $\delta t$ the magnetic field $B'_i$ decreases linearly with time $t$ (i.e., $B'_i \propto t^{-1}$). Actually, the evolution of the magnetic field is caused by the expansion of the shocked regions, which is presented in Figure~\ref{f1}. After the shock crossing time $t_{crs}$ (here, $t_{crs}$ is comparable with the peak time of the slow pulse in GRBs), the spreading of the hot materials
into the vacuum cannot be ignored and the merged shell undergoes
an adiabatic cooling. During this phase, the
volume of the merged shell is assumed to expand as $V '_i \propto R ^s $,
where $s$ is a free parameter and its value is taken to be from 2 to
3. As a result, the particle number density would decrease as
$n'_i \propto {V '}_i^{-1} \propto R ^{-s}$, the internal energy
density as $e '_i \propto {V '}_i^{-4/3} \propto R ^{-4s/3}$, and
the magnetic field strength as $B'_i \propto (e '_i)^{-1/2} \propto
R ^{-2s/3} \propto t ^{-2s/3}$. Because no additional shock-accelerated electrons are injected after the shock crossing time $t_{crs}$,
we only study the prompt emission before $t_{crs}$ in the remaining part of this paper. What we want to point out is that the redefined time interval $\delta t$ is not equal to the shock crossing time ($t_{crs}$), the latter one is dependent on the thickness of the shells. In this paper, the two shells are must be thick enough so that $t_{crs}\gg \delta t$.

The electrons accelerated by the shocks are assumed to have a
power-law energy distribution, $d{N_{e,i}}/d{\gamma '_{e,i}}
\propto {\gamma '_e}^{ - p}$ for ${\gamma '_{e,i}} \ge {\gamma
'_{e,m,i}}$, where ${\gamma '_{e,m,i}}$ is the minimum Lorentz
factor of the accelerated electrons. The following electron cooling discussion is not based on the conventional synchrotron and SSC cooling, which always give us the electron distribution, $\frac{{d{N_e}}}{{d\gamma '_e}} \propto {{\gamma '_e}^{ - 2}}$ for $\gamma'_e < \gamma'_{e,m}$ and $\frac{{d{N_e}}}{{d\gamma '_e}} \propto {{\gamma '_e}^{ - p-1}}$ for $\gamma'_e > \gamma'_{e,m}$ in the fast cooling case, $\frac{{d{N_e}}}{{d\gamma '_e}} \propto {{\gamma '_e}^{ - p}}$ for $\gamma'_e < \gamma'_{e,m}$ and $\frac{{d{N_e}}}{{d\gamma '_e}} \propto {{\gamma '_e}^{ - p-1}}$ for $\gamma'_e > \gamma'_{e,m}$ in the slow cooling case. These electron distributions do not take into account the evolution of the magnetic field. Ref.~\cite{uhm14}  discussed the electron distribution affected by a decaying magnetic field based on $B' \propto r^{-b}$, where $r$ is the fireball radius and $b$ is the magnetic field decaying index. They considered the electron distribution of a group of plasma in a magnetic field with an arbitrary decaying index $b$, which is called a ``toy box model". Here we consider a more physical process, internal shocks, which generate an evolutional magnetic field and a consistent spectrum with the observed Band spectral shape.

In the comoving frame, the evolution of the Lorentz factor of an electron via synchrotron and SSC cooling and adiabatic cooling can be described by~\cite{uhm14}
\begin{equation}
\frac{d}{{dt'}}\left(\frac{1}{{{\gamma _e}}}\right) = \frac{{{\sigma _T}(1+Y_i)}}{{6\pi {m_e}c}}{{B'_i}^2} - \frac{1}{3}\left(\frac{1}{{{\gamma _e}}}\right)\frac{{d\ln n'_i}}{{dt'}},
\label{gammacooling}
\end{equation}
where $Y_i\approx[(4\eta_{i}\epsilon_e/\epsilon_B+1)^{1/2}-1]/2$ is the
Compton parameter, which is defined by the ratio of the IC to
synchrotron luminosity, with $\eta_{i}=\min[1,({\gamma
'_{e,c,i}}/{\gamma '_{e,m,i}})^{2-p}]$~\citep{sar01}. $\gamma
'_{e,c,i}$ is the cooling Lorentz factor and the comoving time $t'=2 \gamma t$.

The minimum Lorentz factor of the accelerated electrons is ${\gamma '_{e,m,i}} =
\frac{{{m_p}}}{{{m_e}}}(\frac{{p - 2}}{{p - 1}}){\epsilon
_e}({\gamma _{\rm rel}} - 1)$ (where ${\gamma _{\rm
rel}}=\gamma_{21}$ or $\gamma_{34}$ in region 2 or 3), so it can be written as:
\begin{equation}
\begin{array}{c}
{\gamma '_{e,m,3}}
 \simeq 1.0 \times {10^4}{g_{p}}{\epsilon _{e, - 1/2}}{\gamma _{4,4.5}}\gamma _{1,2.5}^{ - 1},
\end{array}
\end{equation}
\begin{equation}
\begin{array}{c}
{\gamma '_{e,m,2}}
 \simeq 30{g_{p}}{\epsilon _{e, - 1/2}},
\end{array}
\end{equation}
\textls[-45]{where $g_p=3(p-2)/(p-1)$.
Moreover, the cooling Lorentz factor ${\gamma '_{e,c,i}} = 6\pi
{m_e}c/({y_i}{\sigma _T}{B'_3}^2\gamma t)$, can be written as}
\begin{equation}
{\gamma '_{e,c,3}} = {\gamma '_{e,c,2}} \simeq 3.4 \times {10^2}
y_{i,0}^{ - 1}\epsilon _{B, - 1/2}^{ - 1}L_{k,51}^{ - 1}\gamma _{1,2.5}^5
\frac{{(t + \delta t)_{,0}^2}}{{{t_{, 0}}}},
\end{equation}
where $y_i=1+Y_i$ is the ratio of the total luminosity to
synchrotron luminosity.

\begin{figure}[H]
 \includegraphics[width=10.5 cm]{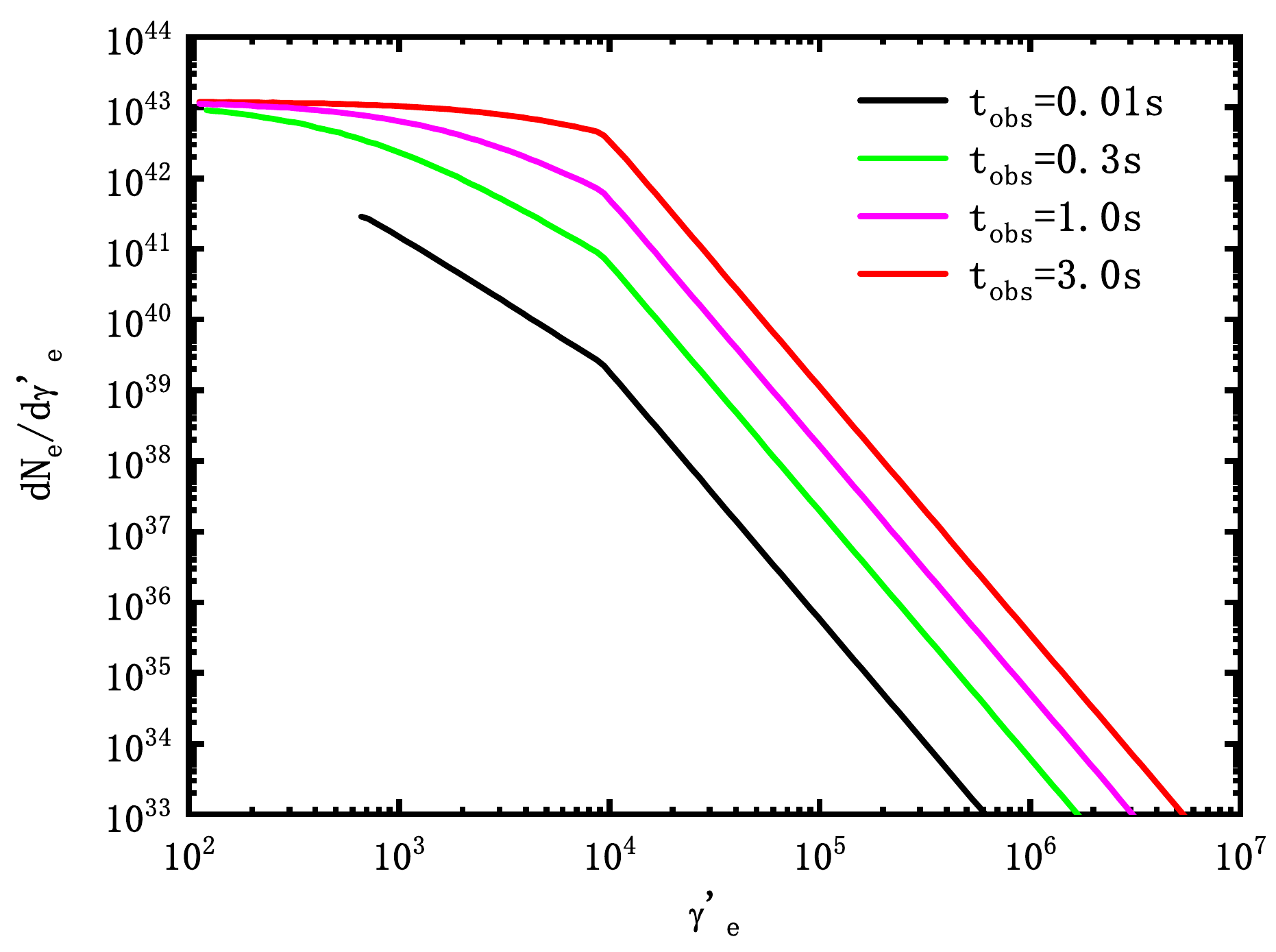}
  \caption{The electron distribution in energy space after cooling time $t$ in the evolutional magnetic field in Figure~\ref{f1}. The same $\delta t$, $t_{crs}$, and dynamics parameters as in Figure~ \ref{f1} are taken in numerical calculations.\label{f2}} %MDPI please put after First mentioned.
\end{figure}

From the electron injection rate based on  {Equations~(\ref{dNe2}) and (\ref{dNe3})}, one can obtain the injected electrons number between $t'$ and $t'+dt'$. Assuming the original electron injection distribution $d{N_{e,i}}/d{\gamma '_{e,i}} \propto {\gamma '_e}^{ - p}$ for ${\gamma '_{e,i}} \ge {\gamma'_{e,m,i}}$, the injected electrons number between $t'$ and $t'+dt'$ and between $\gamma'_e$ and $\gamma'_e+d\gamma'_e$ can be derived. So we cut the injected electrons into small pieces in the time space $t'$ and the energy space $\gamma'_e$. At the beginning, time $t'=0$, a number of electrons $dN$ will be injected into the shocked region in a time interval $dt'$ and will be cooled in the initial magnetic field, so one can obtain the change of electron Lorentz factor $\Delta \gamma'_{e,1}$ based on  {Equation~(\ref{gammacooling})} for the electrons between $\gamma'_e$ and $\gamma'_e+d\gamma'_e$. In the next time interval $dt'$, these electrons with the Lorentz factor between $\gamma'_e+\Delta \gamma'_{e,1}$ and $\gamma'_e+d\gamma'_e+\Delta \gamma'_{e,1}$ will be cooled in the instantaneous magnetic field based on the evolutional magnetic field in {Equation~(\ref{b})}, and one can obtain another $\Delta \gamma'_{e,2}$ ($\Delta \gamma'_{e,2} \ne \Delta \gamma'_{e,1}$). At the same time, another group electrons are injected and cooled in this instantaneous magnetic field. These processes are continuous before $2\gamma t_{crs}$. The shocked electrons are injected as time and all the electrons are cooled in the instantaneous magnetic field. We sum all electrons at time $t'$ in the energy space, obtain the electron distributions at time $t'$ and present them in Figure~\ref{f2} ($t_{obs}=t'/2\gamma$). As shown in Figure \ref{f2}, when $t < \delta t$, the magnetic field does not change significantly (see Figure~\ref{f1}), the electron distribution in the fast cooling case, $\frac{{d{N_e}}}{{d\gamma '_e}} \propto {{\gamma '_e}^{ - 2}}$ for $\gamma'_e < \gamma'_{e,m}$, and $\frac{{d{N_e}}}{{d\gamma '_e}} \propto {{\gamma '_e}^{ - p-1}}$ for $\gamma'_e > \gamma'_{e,m}$, are expected. However, when $t > \delta t$, the electron distribution below $\gamma'_{e,m}$ would be flattened because of the decaying magnetic field. Due to the magnetic field decay, the electrons injected at later times would cool more slowly than the electrons injected at early times (here, all times are before $t_{crs}$). In other words, the cooling efficiency would become smaller due to the decaying magnetic field, which induces more electrons accumulating at $\lesssim \gamma'_{e,m}$ than in the invariable magnetic field case. When $t\gg \delta t$ and $t<t_{crs}$, the electron spectral index for $\gamma'_e < \gamma'_{e,m}$ is even flattened to~zero.

In order to find these new results for $t\gg \delta t$ and $t<t_{crs}$, we can evaluate the continuity equation of electrons in energy space, $ \frac{\partial }{{\partial t'}}{({d{N_{e,\gamma'_e }}}}/{{d\gamma '_e})} + \frac{\partial }{{\partial {\gamma' _e}}}[\dot {\gamma'_e} {({d{N_{e,\gamma'_e }}}}/{{d\gamma '_e}})] = Q(\gamma'_e ,t') $, where ${d{N_{e,\gamma'_e }}}/{{d\gamma '_e}}$ is the instantaneous electron spectrum at time $t'$, and $Q(\gamma'_e ,t')=Q_0 (t') (\gamma'_e/\gamma'_{e,m})^{-p}$ is the electron injection distribution accelerated by shocks above the minimum injection Lorentz factor $\gamma'_{e,m}$. By ignoring the inconsequential adiabatic cooling term, we can get $\frac{d}{{dt'}}(\frac{1}{{{\gamma '_e}}}) \propto (1 + {Y_i}){B'_i}^2 \propto {{t'}^{ - 2}} $, where $Y_i$ is assumed to be a constant before the shock crossing time $t_{crs}$ in the fast cooling case. Then, we can obtain $\gamma'_e \propto {t'}$, and thus $ {{\dot \gamma '}_e} \propto {\gamma '_e}^2 {{t'}^{ - 2}} \propto {\gamma '_e}^0 $.
For $\gamma'_{e,c} < \gamma'_e < \gamma'_{e,m}$, $Q(\gamma'_e ,t')=0$, to obtain the final and quasi-steady electron spectral shape at the arbitrary time $t'$, by considering a quasi-steady-state system ($\partial /\partial t = 0$), we can easily find ${d{N_{e,\gamma'_e }}}/{{d\gamma '_e}}\propto {\gamma '_e}^0 $ below $\gamma'_{e,m}$.

Next, the four characteristic frequencies in regions 2 and 3 that can be calculated from ${\nu} = \frac{{{q_e}}}{{2\pi {m_e}c}}{B'}{\gamma '_e} ^2 \gamma$ are derived as~\cite{wang13}
\begin{equation}
{h\nu _{m,2}} \simeq 2.1 \times {10^{-4}} g_{p}^2\epsilon _{e, -
1/2}^2\epsilon _{B, - 1/2}^{1/2}L_{k,51}^{1/2}\gamma _{1,2.5}^{ - 2}(t
+ \delta t)_{,0}^{ - 1} \; \mathrm{keV},\label{eqnum2}
\end{equation}
\begin{equation}
{h\nu _{m,3}} \simeq 26  g_{p}^2\epsilon _{e, -
1/2}^2\epsilon _{B, - 1/2}^{1/2}L_{k,51}^{1/2}\gamma
_{4,4.5}^2\gamma _{1,2.5}^{ - 4}(t + \delta t)_{,0}^{ - 1} \; \mathrm{keV},\label{eqnum3}
\end{equation}
and
\begin{equation}
{h\nu _{c,2}}={h\nu _{c,3}} \simeq 3.6 \times {10^{-2}}y_{,0}^{ -
2}\epsilon _{B, - 1/2}^{ - 3/2}L_{k,51}^{ - 3/2}\gamma
_{1,2.5}^8\frac{{(t + \delta t)_{,0}^3}}{{t_{, 0}^2}} \; \mathrm{keV}.\label{eqnuc}
\end{equation} %MDPI please confirm whether need indent.
Here, if $\gamma _1 =100$ and $\gamma _4 =10,000$, we obtain $h\nu _{m,3}\simeq 186\; \mathrm{keV}$ at time $t=1\; \mathrm{s}$, which is approximatively equal to the typical value of $E_p$ of the GRB prompt emission.

\begin{figure}[H]
 \includegraphics[width=10.5 cm]{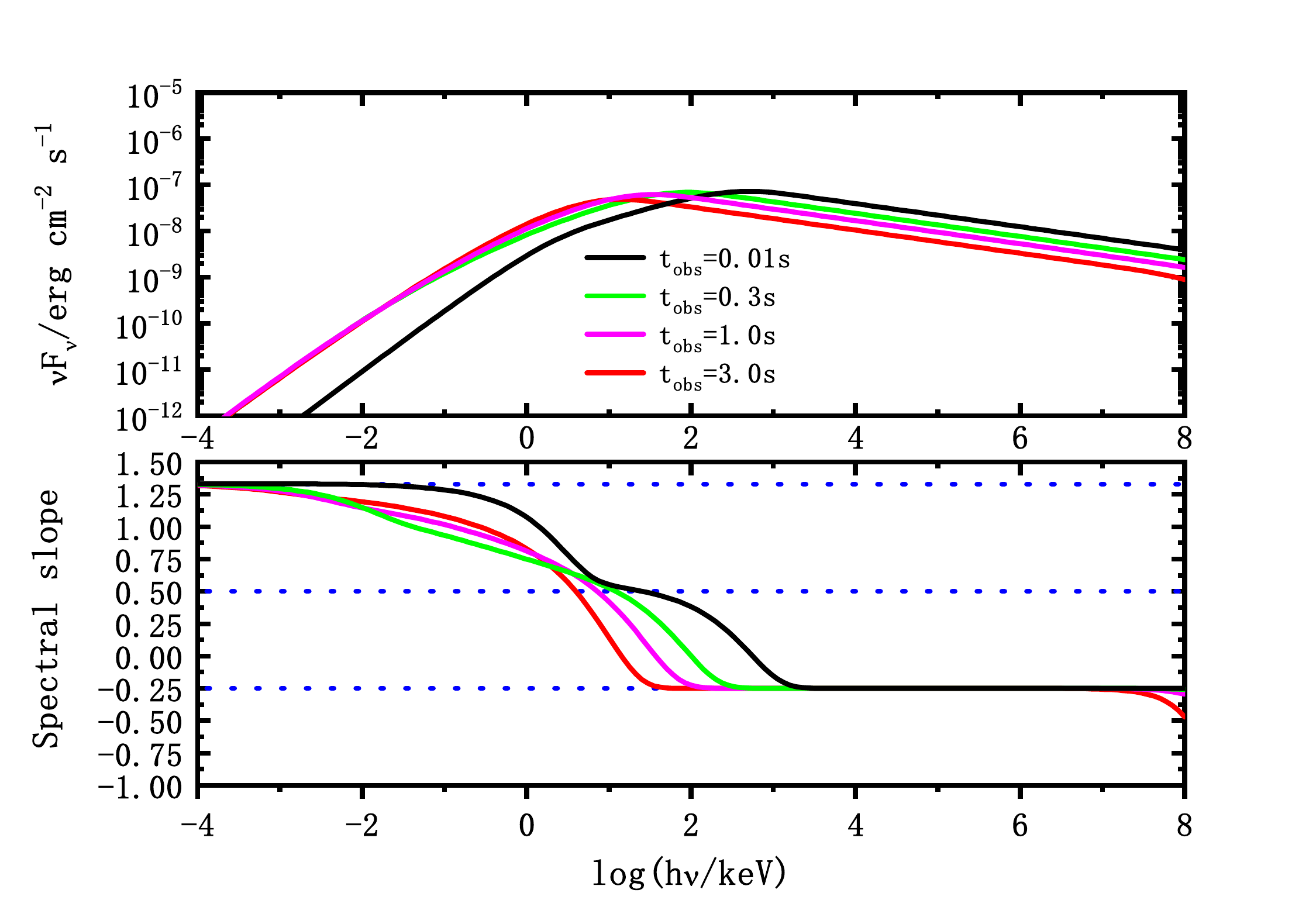}
  \caption{The top panel corresponds to time-resolved spectra in four different $t$ as in Figure~\ref{f2} and the bottom panel shows corresponding synchrotron spectral slopes. The same $\delta t$, $t_{crs}$, and dynamics parameters as in Figure~\ref{f1} are taken in numerical calculations.\label{f3}}
\end{figure}

We also present the spectrum of region 3 in the top panel of Figure~\ref{f3} based on the electron distribution shown in Figure~\ref{f2}. However, we do not present the spectrum of region 2 because, from NFS, (1) its photon peak frequency is much smaller than the typical GRB prompt emission $E_p$, (2) the radiation efficiency can not be high enough as a result of slow cooling, and (3) the flux of region 2 is much lower than that of region 3. The last reason can be evaluated from~\citep{sar98,yu09}
\begin{equation}
{F_{\nu ,\max ,i}} < \frac{{{N_{e,i}}}}{{4\pi D_L^2}}\frac{{{m_e}{c^2}{\sigma _T}}}{{3{q_e}}}{B'_i}\gamma , \label{fnumaxi}
\end{equation}
where $D_{L}$ is the luminosity distance of the burst and $N_{e,i}$ is the total number of injected electrons until the time $t$. Since a portion of the electrons have cooled to a much smaller value than $\gamma_{break}$ (where the break Lorentz factor $\gamma_{break}$ of an electron distribution, $\gamma_{break}=\gamma_{e,m}$ for fast cooling, and $\gamma_{break}=\gamma_{e,c}$ for slow cooling), the actual number of electrons near $\gamma_{break}$ is less than $N_{e,i}$ and thus the actual ${F_{\nu ,\max ,i}}$ is smaller than the right term of inequality Equation~(\ref{fnumaxi}). So, we can obtain
\begin{align}
{\nu _{m,3}}{F_{\nu ,\max ,3}} < & 1.5 \times {10^{  5}}g_p^2\epsilon _{e, - 1/2}^2{\epsilon _{B, - 1/2}}L_{k,51}^2 {\gamma _{4,4}}\gamma _{1,2}^{ - 6}\nonumber \\
& \times \frac{{{t_{,0}}}}{{(t + \delta t)_{,0}^2}}D_{L,28}^{ - 2} \; \mathrm{keV \, cm^{-2} s^{-1}} ,
\label{nufnumax}
\end{align}
and
\begin{align}
{\nu _{c,2}}{F_{\nu ,\max ,2}} < & 8.7 \times {10^{ - 1}}y_{,0}^{ - 2}\epsilon _{B - 1/2}^{ - 1}\gamma _{1,2}^5 \nonumber \\
& \times \frac{{(t + \delta t)_{,0}^2}}{{{t_{,0}}}}D_{L,28}^{ - 2} \; \mathrm{keV \, cm^{-2} s^{-1}},
\end{align}
where $\gamma_1=100$ and $\gamma_4$ = 10,000  are taken.

From Figure~\ref{f3}, we can see that for $t<\delta t$, because of a constant magnetic field, the spectral slope of $\nu F_{\nu}$ is $1/2$ as described by~\cite{sar98}. However, when $t>\delta t$, the spectral slope will deviate from $1/2$ and become a larger value (even $4/3$). If the electron index ($d{N_e}/d{\gamma _e} \propto \gamma _e^{ - u}$) is $u$ , the $F_\nu$ slope of synchrotron radiation (${F_\nu } \propto {\nu ^{ - w}}$) would be $w = (u - 1)/2$  and the photon spectral index (defined as $d{N_\gamma }/d{E_\gamma } = E_\gamma ^{ - \alpha }$, where $E_\gamma$ is the photon energy, and $N_\gamma$ is the photon number flux) would be $\alpha  =  - (w + 1)$. Due to the decaying magnetic field, $u$ tends to be zero, and thus $w=-1/2$ and $\alpha = -1/2$. However, when $\alpha>-2/3$, because of the overlying effect, the low energy photon index of the electrons with $\sim \gamma_{e,m}$ is $-2/3$ and will cover the emission of electrons with smaller Lorentz factors. So, due to the effect of the low-energy radiation tail of electrons with Lorentz factor $\gamma_{e,m}$, $\alpha$ is at most equal to $-2/3$, and we can get $-3/2<\alpha<-2/3$. This is consistent with the observations~\citep{pre00,gru14}, which suggest $\alpha  \sim  -1$.

\subsection{The Effect of the Variable Electron Injection Rate}\label{sec32}

Although theoretically the low-energy photon spectral index $\alpha$ can reach $-2/3$ caused by a decaying magnetic field, since this is a gradual process, $\alpha$ would be softer than $-2/3$ for $E \lesssim E_p$, which can be seen from Figure~\ref{f3}. In fact, $\alpha \sim -1$ can be fitted easily, but fitting $\alpha$ slightly smaller than $-2/3$ is difficult. We here consider a variable electron injection rate, which could induce $\alpha \sim -2/3$. The variable electron injection rate may be suggested by that the actual GRB shell is not homogeneous and presents a density profile, for example, a Gaussian density profile, inducing a rising electron injection rate. Nonetheless, we do not know its growing method clearly. Ref.~\citep{uhm14}  discussed this effect in their ``toy box model'', and suggested, because of a rising electron injection rate, $\alpha$ goes from $-0.82$ to $-1.03$, which is dependent on the growing power-law index $q$ (where the injection rate $\propto {t'} ^q$ with $q=1$, $2$ or $3$). Here we adopt similar expressions of the rising electron injection rate,
\begin{equation}
{dN_{e,2}} = 8\pi {R^2}{n'_1}({\gamma _{21}}{\beta _{21}}/\gamma
\beta ){\gamma ^2}c dt \times {\left(\frac{t}{{{t_0}}}\right)^q}
\label{Ne2}
\end{equation}
and
\begin{equation}
{dN_{e,3}} = 8\pi {R^2}{n'_4}({\gamma
_{34}}{\beta _{34}}/\gamma \beta ){\gamma ^2}c dt  \times {\left(\frac{t}{{{t_0}}}\right)^q},
\label{Ne3q}
\end{equation}
where the factor $(\frac{t}{{{t_0}}})$ is to maintain the same electron injection number in the interval $t_0$ as that for the constant injection rate $q=0$.

\begin{figure}[H]
  \includegraphics[width=10.5 cm]{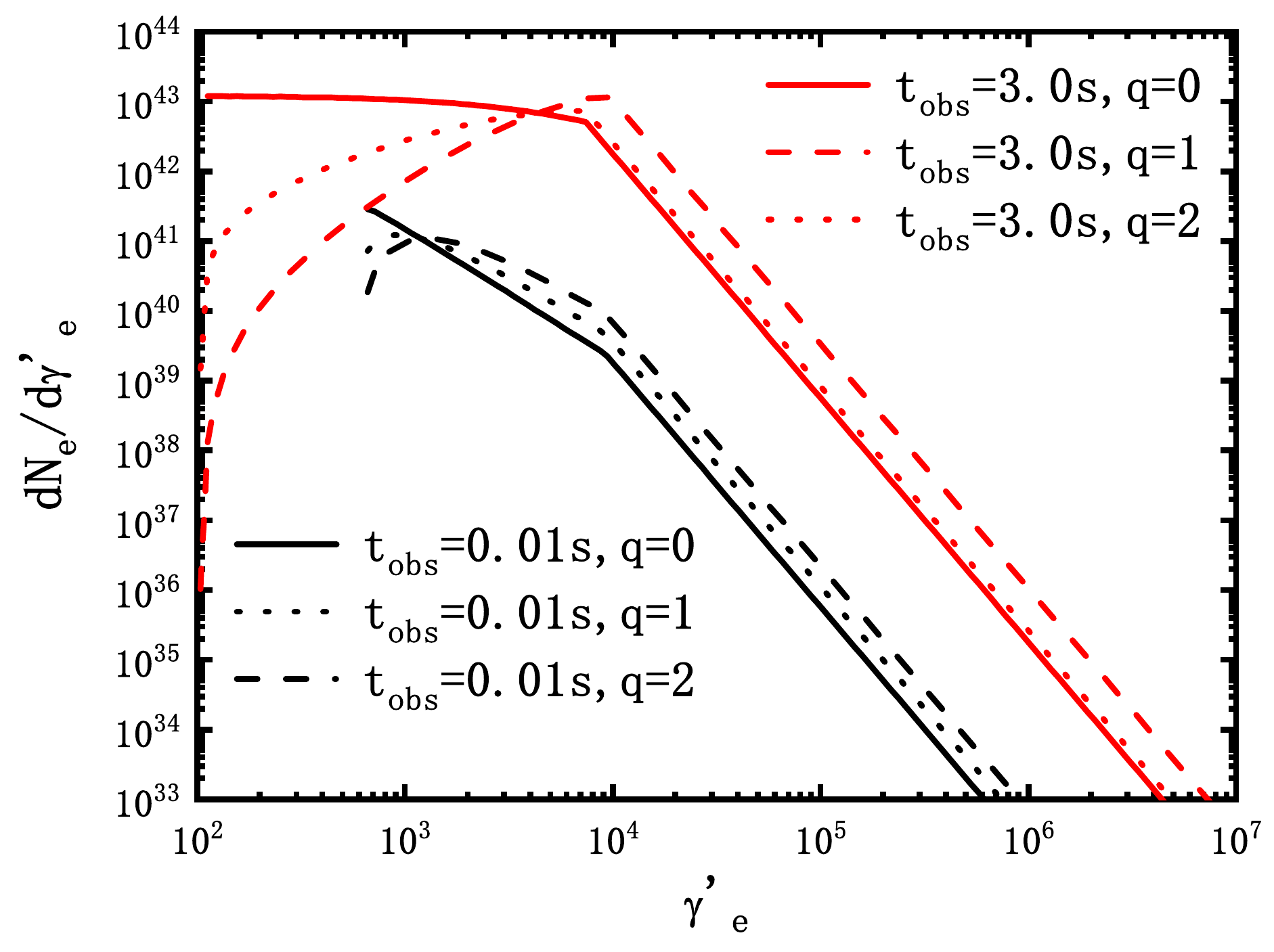}  

    \includegraphics[width=10.5 cm]{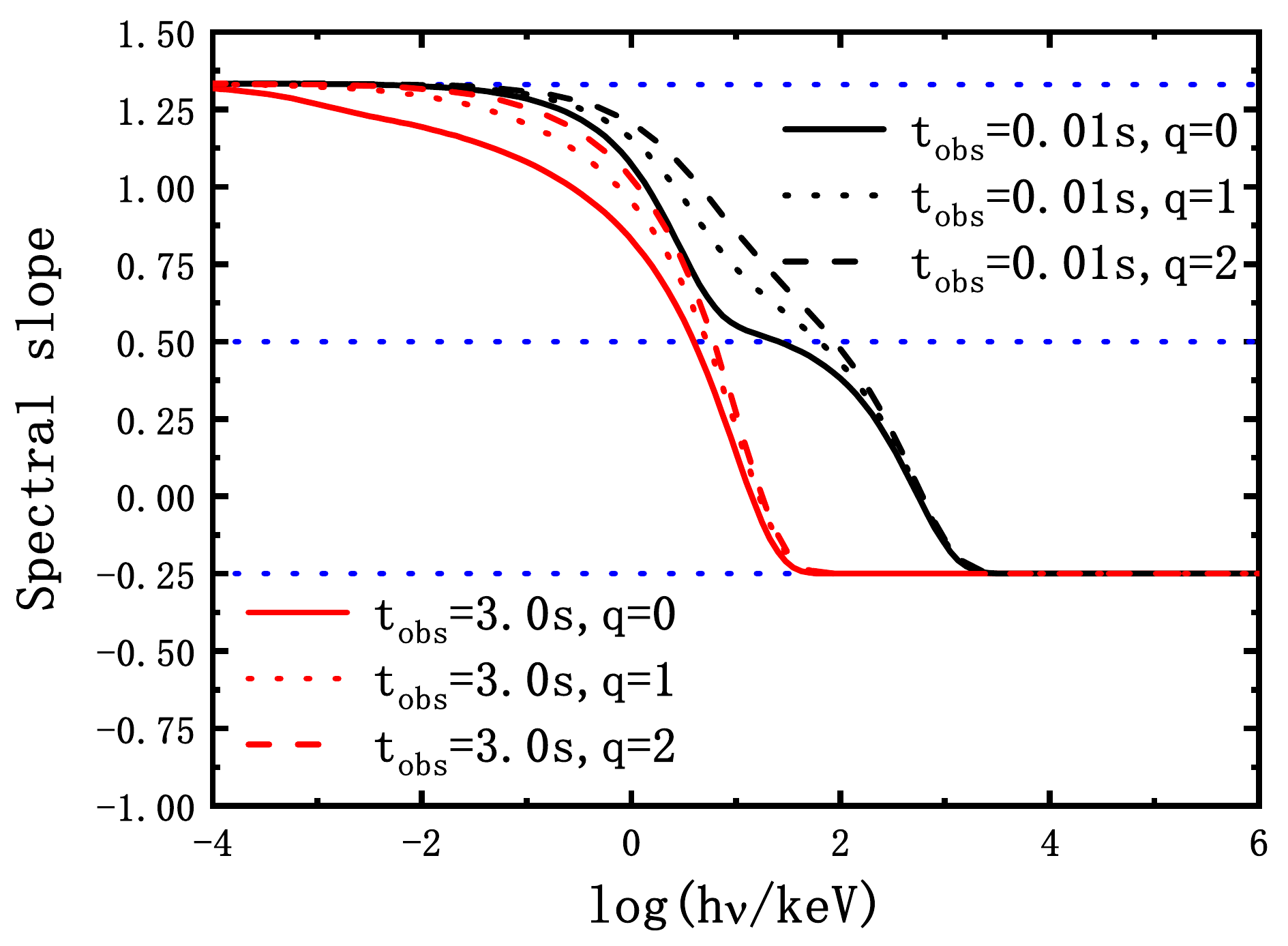}
  \caption{The top panel shows the electron distributions in evolutional magnetic fields and different electron injection rising indices. We adopt the electron injection rising index, $q=0$ (solid line), $q=1$ (dotted line) and $q=2$ (dashed line). The bottom panel shows the corresponding synchrotron spectral slopes for these electron distributions. The same parameters as in Figure~\ref{f1} are taken for numerical calculations.\label{f56}}
\end{figure}

We show the electron distribution for a rising electron injection rate in the top panel of Figure~\ref{f56}. The rising electron injection would increase the electrons injected later, which would cool in a weaker magnetic field and pile up at $\lesssim \gamma'_{e,m}$. This can result in a harder electron distribution and a relative spectrum. The slopes of the spectra are presented in the bottom panel of Figure~\ref{f56}. We can see that the slopes of the spectra tend to reach $-4/3$ more easily than in the constant electron injection case. In addition, a larger $q$ would generate a harder low-energy spectral index. \vspace{-6pt}

\section{Application to the Actual GRB Spectra}

\begin{figure}[H]
  \includegraphics[width=10.5 cm]{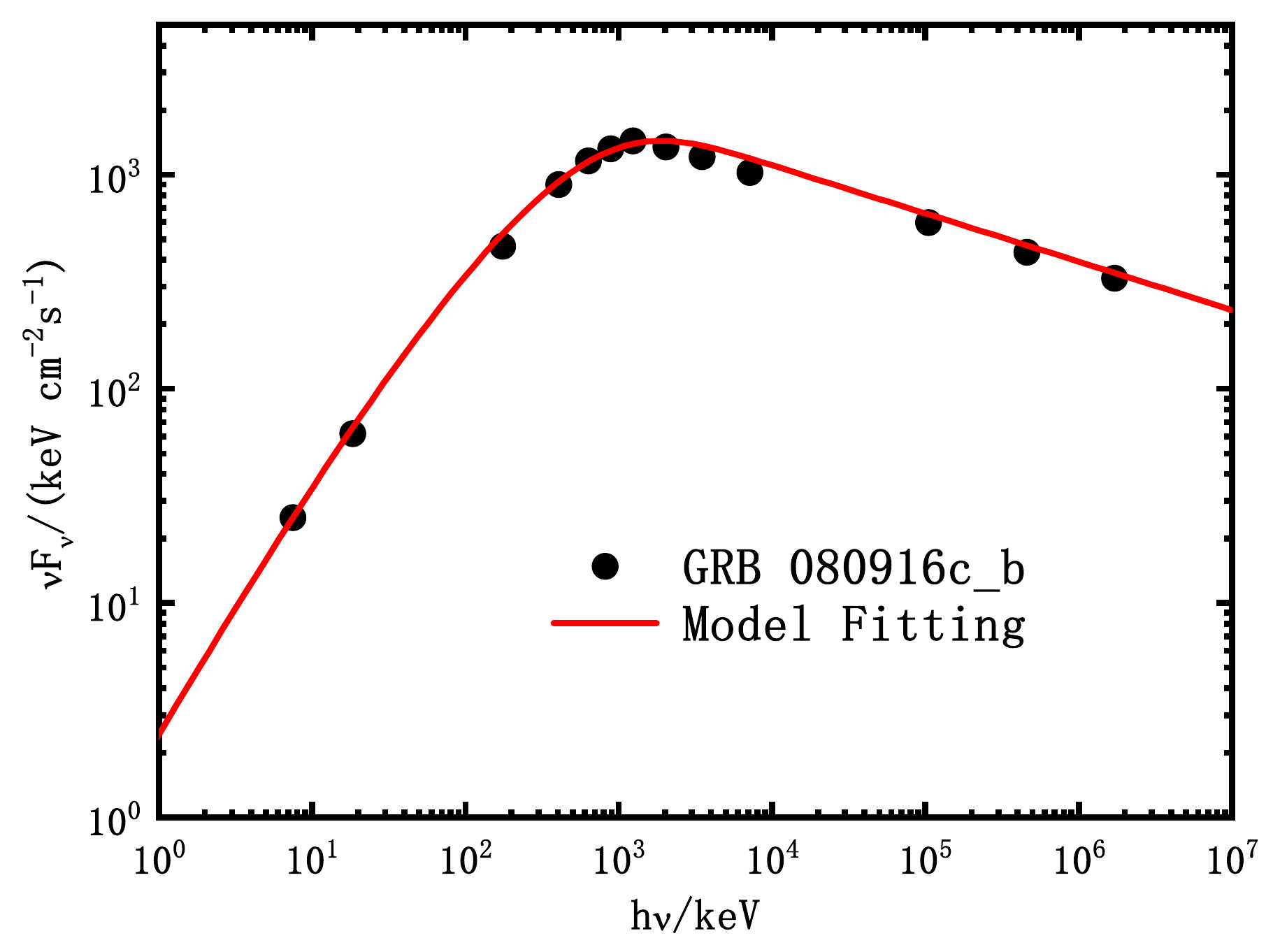}
  \caption{The time-averaged spectrum to fit the interval ``b" of GRB 080916c. The black points are selected from the Band function with the low-energy photon index $\alpha=-1.02\pm0.02$, the high-energy photon index $\beta=-2.21\pm0.03$ and the peak energy $E_p=1170\pm140\,\mathrm{keV}$ provided in Ref.~\cite{abdo09a}, which are precise enough to present the tendency of the actual GRB emission. This spectral duration is from $3.58\;\mathrm{s}$ to $7.68\;\mathrm{s}$ and we fit it by adopting the time-averaged spectrum from $t =0 \; \mathrm{s}$ to $t=3 \; \mathrm{s}$ ($t_{max} \le t_{crs}$). The fitting parameters are listed in Table~\ref{tab1}.\label{f4}}
\end{figure}

In order to compare with the actual GRB spectrum, we select the broad band spectrum of GRB 080916c in the interval ``b" detected by Gamma-ray Burst Monitor (GBM) and the Large Area Telescope (LAT) aboard the Fermi satellite (see Ref.~\cite{abdo09a}), from $3.58\,\mathrm{s}$ to $7.68\,\mathrm{s}$ since the lightcurve during this period is presented as a single and pure pulse. Moreover, its low energy photon index is close to the typical value of the low energy photon index of the GRB, that is, $\alpha \sim -1$, harder than the expectation of synchrotron fast cooling ($-1.5$). The ``b" spectrum of GRB 080916c can be well fitted by the Band function with the low energy photon index of $\alpha=-1.02\pm0.02$, the high energy photon spectral slope $\beta=-2.21\pm0.03$ and the peak energy $E_p=1170\pm140\,\mathrm{keV}$~\citep{abdo09a}. Since the observational data can be well fitted by this Band function with a very small error range, the Band function is precise enough to represent the actual GRB emission. We select some representative points (black points in Figure~\ref{f4}) in this Band function to present the tendency of the actual GRB emission. In addition, more black points around the peak energy in the figure are taken to present the gradual change in behavior there. In Figure~\ref{f4}, by using a time-averaged energy spectrum from $t=0\,\mathrm{s}$ to $t=3\,\mathrm{s}$, the emission of GRB 080916c can be fitted well in our model with the proper parameters, which have been listed in Table~\ref{tab1}.

\begin{figure}[H]
  \includegraphics[width=10.5 cm]{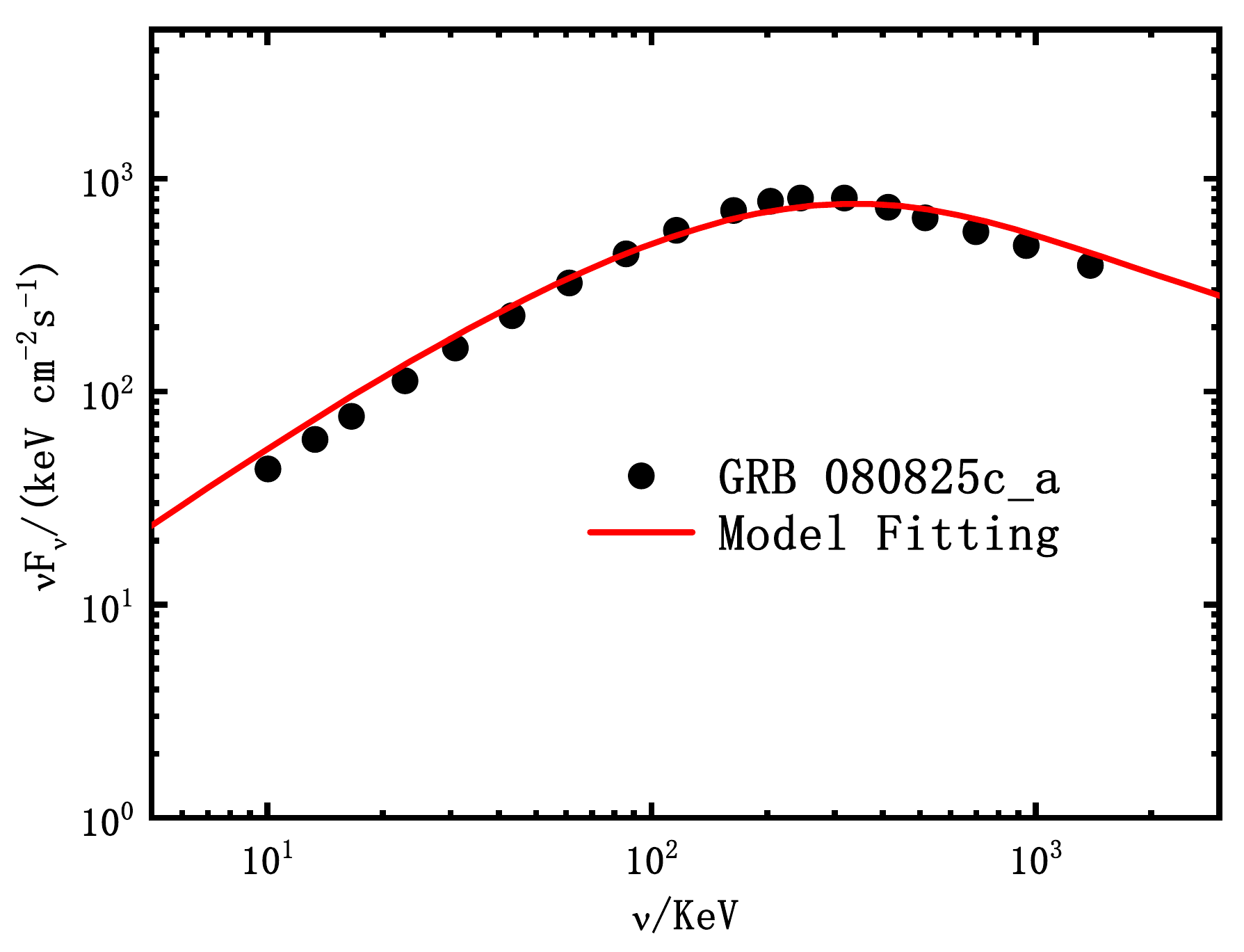}
  \caption{The time-averaged spectrum to fit the interval ``a'' of GRB 080825c. The black points are selected from the Band function with the low-energy photon index $\alpha=-0.76 \pm 0.05$, the high-energy photon index $\beta=-2.54_{-0.17}^{+0.11}$ and the peak energy $E_p=291_{-22}^{+25}\,\mathrm{keV}$ provided in Ref.~\cite{abdo09b}, which are precise enough to present the tendency of the actual GRB emission. This spectral duration is from $0.0\,\mathrm{s}$ to $2.7\,\mathrm{s}$ and we fit it by adopting the time-averaged spectrum from $t =0 \; \mathrm{s}$ to $t=3 \; \mathrm{s}$ ($t_{max} \le t_{crs}$). The fitting parameters are listed in Table~\ref{tab1}.\label{f7}}
\end{figure}

We select the single pulse spectrum of GRB 080825c in the interval ``a'' detected by Fermi GBM and LAT (see Ref.~\cite{abdo09b}), from $0.0\,\mathrm{s}$ to $2.7\,\mathrm{s}$, which has a harder photon index, $\alpha\sim-0.76 $. The ``a''  spectrum of GRB 080825c can be well fitted by the Band function with the low energy photon index $\alpha=-0.76 \pm 0.05$, the high energy photon index $\beta=-2.54_{-0.17}^{+0.11}$ and the peak energy $E_p=291_{-22}^{+25}\,\mathrm{keV}$ ~\citep{abdo09b}. Such a hard photon index could not be approached easily for a constant electron injection rate, that is, $q=0$, so we consider a rising electron injection rate as suggested in Section~\ref{sec32}. Some representative points (black points in Figure~\ref{f7}) in this Band function are selected to present the tendency of the actual GRB emission as the same as the treatment for the GRB 080916c. The observational spectrum can be reproduced well in Figure~\ref{f7} phenomenally by using a time-averaged energy spectrum from $t=0\,\mathrm{s}$ to $t=3\,\mathrm{s}$ with an index of rising electron injection rate $q=2$ and other reasonable parameters (all parameters are listed in Table~\ref{tab1}).

\unskip
\begin{specialtable}[H] 
\setlength{\tabcolsep}{4.5mm}  
\caption{The parameters adopted  to fit the spectra of GRB 080916c and GRB 080825c.\label{tab1}}
%%% \tablesize{} %% You can specify the fontsize here, e.g., \tablesize{\footnotesize}. If commented out \small will be used.
\begin{tabular}{cccc}
\toprule
\textbf{Parameters}	& \textbf{Symbol} & \textbf{GRB 080916c}	& \textbf{GRB 080825c}\\
\midrule
Redshift		& $z$			& 4.35     & 1\\
Index of electron injection rate		& $q$			& 0     & 2\\
redefined time interval	(s)	& $\delta t$			& 0.1     & 0.1\\
Shock cross time (s)		& $t_{crs}$			& 3      & 3\\
Kenetic luminosity (erg/s)		& $L_{k}$			& 3.3$ \times 10^{53}$      & 1.2$ \times 10^{51}$\\
Bulk Lorentz factor of region 1		& $\gamma_{1}$			& 146      & 255\\
Bulk Lorentz factor of region 2		& $\gamma_{4}$			& $3 \times 10^{4}$      & $3 \times 10^{4}$\\
Electron injection index		& $p$			& 2.5      & 3.2\\
Electron equipartition factor		& $\epsilon_e$			& 0.3      & 0.3\\
Magnetic equipartition factor		& $\epsilon_B$			& 0.3      & 0.3\\
\bottomrule
\end{tabular}
\end{specialtable}

\textls[-15]{The main parameters to effect the final spectrum are listed in Table~\ref{tab1}. The dependence of the break energy of the spectrum on the listed parameters could be found in Equation~(\ref{eqnum3}) and for the magnitude of peak flux the dependence could be derived roughly in Equation~(\ref{nufnumax}).} During the model fitting, for simplification, the energy equipartition factors for electrons and the magnetic field, that is,$\epsilon_e$ and $\epsilon_B$, and $\gamma_4$ are fixed, and then the Lorentz factor $\gamma_1$ and kinetic luminosity $L_k$ are adjusted to match the observational peak energy and the peak flux. The electron injection index $p$ is determined by the observational high-energy photon index since the relation between them is $\beta \sim (-p-2)/2$, \endnote{{The high-energy electron} distribution above the break electron energy is $d{N_e}/d{\gamma _e} \propto  {\gamma _e}^{(-p-1)}$.}  %MDPI not allowed footnote, please check and modify. Should as ``'Notes', before reference part.
suggested by the synchrotron radiation. The shock cross time is comparably adopted with the typical duration of the slow pulse of the GRB, namely, $\sim 3 \,\rm s$. Different values of $\delta t$ could affect the evolutional form of the magnetic field (as shown in Figure~\ref{f1}) and adjust the weight of the cooling in a constant magnetic field and the cooling in
a decaying magnetic field. In other words, a smaller $\delta t$ could make it so that the electron synchroton cooling mainly takes place in a decaying magnetic field and leads the photon index to be harder, while for a larger $\delta t$, the electrons are mainly cooling in a constant magnetic field and generating a photon index close to $-1.5$. As a result, for GRB 080916c with a photon index $\sim -1$, a relatively small $\delta t=0.1\,\rm s$ is adopted. A harder photon index $\sim -0.76$ for GRB 080825c and a rising electron injection rate with an index $q=2$ are taken into account as suggested in Section~\ref{sec32}. Therefore, a certain range of a low-energy photon index from $\sim -3/2$ to $\sim -2/3$ could be approached through the adjustment of $\delta t$ and the index of the electron injection rate $q$. However, for a low-energy photon index harder than $ -2/3$, this model would become invalid.

\section{Discussions and Conclusions}
 Currently alleviating the tension between the expectation of synchrotron and observations in the GRB prompt regime is a more and more important issue. Two classes of model have been proposed to explain the low-energy photon index of GRB prompt emission, Comptonized quasi-thermal emission from the photosphere within a relativistic outflow and synchrotron and/or SSC emission in the optically thin region. These models can experience difficulties. For Comptonized quasi-thermal emission, the most significant effect to obtain $\alpha \sim -1$ is the equal arrival time effect in this model, which is relevant to the end time of central engine activity, but may not be applicable during the prompt emission phase when a continuous wind is ejected from the central engine~\citep{zhang11b}. Synchrotron slow cooling in internal shocks may not provide a high radiative efficiency, and a dominant SSC component usually predicts an even more dominant 2nd-order SSC component, which significantly exceeds the total energy budget of GRBs~\citep{der01,piran09}. Thus, some evolutional parameters, such as the magnetic field, the fraction of the accelerated electrons, and the energy equipartition factors, were suggested to explain the low-energy index.

In this paper, we have considered a straightforward model, that is, the fast cooling synchrotron radiation in internal shocks. We obtain the magnetic field evolutional form in a practical shell--shell collision, $B' \propto $ constant before $\delta t$ and $B' \propto t^{-1}$ after this time, and recalculate the electron distribution for this evolutional magnetic field. When $t < \delta t$, the magnetic field is nearly constant, and the fraction of cooling electrons in the invariable magnetic field is high enough so that $\frac{{d{N_e}}}{{d\gamma '_e}} \propto {{\gamma '_e}^{ - 2}}$ for $\gamma'_e < \gamma'_{e,m}$ is expected. However, when $t \gg \delta t$ but $t < t_{crs}$, the fraction of cooling electrons in the evolutional magnetic field is higher than in the invariable magnetic field, so that $\frac{{d{N_e}}}{{d\gamma '_e}}$ will be gradually proportional to $ {{\gamma '_e}^{ 0}}$. $2\gamma^2 c\delta t$ and $2\gamma^2ct$ indicate roughly the collision radius and the propagation distance of a relativistic outflow after the collision takes place but before the shock crossing time $t_{crs}$, respectively. In other words, if the propagation distance of the outflow is smaller than the collision radius before the shock crossing time, the magnetic field can be treated as a constant and it is not necessary to consider the evolution of the magnetic field when calculating the electron cooling. However, if the propagating distance of the outflow is larger than the collision radius before the shock crossing time, we have to consider the evolution of the magnetic field and can obtain a different electron distribution. Actually, the outflow may undergo the first case and then the second case, so we can obtain a reasonable range of the low-energy photon index $\alpha$, from $-3/2$ to $-2/3$ theoretically. Since $\frac{{d{N_e}}}{{d\gamma '_e}}$ proportional to $ {{\gamma '_e}^{ 0}}$ is a gradual process below $E_p$, it is usually difficult to get $\alpha$ to be exactly equal to $-2/3$, but this index is only slightly smaller than $-2/3$. Moreover, we also consider a rising electron injection rate, which may exacerbate this situation, inducing $\alpha$ to be closer to $-2/3$.

Ref.~\cite{uhm14} considered a decaying magnetic field varying with the distance from the central engine to explore the range of a low energy photon index in the GRB prompt regime. They discussed the radiation spectra of a cloud of plasma in a decaying magnetic field with an arbitrary decaying index $b$ for a simplified model, which is called a ``toy box model''. Different from their work, we adopt a more physical case for the internal shock by considering the collision of two shells and inducing the decaying form of the magnetic field. As a result, a time-dependent magnetic field is derived (as shown in Figure \ref{f1}). In fact, a time-dependent magnetic field could be translated to a distance-dependent form due to the propagation of relativistic outflow. For the evolutional magnetic field form obtained from the practical internal shock, we study the influence on the spectral index. In addition to the detailed treatment of shell--shell collision, the kinetic luminosity and the energy equipartition parameters, $\epsilon_{B}$ and $\epsilon_{e}$ are taken into account to obtain the radiation spectra, comparing them with the actual GRB spectra for GRB 080916c and GRB 080825c.
%In fact, the quasi-thermal component and the non-thermal component is likely to coexist in the prompt emission of GRBs. When the former is dominant, %we tend to see the quasi-thermal component. e.g., GRB 090902b~\citep{abdo09a}. But if the latter is dominant, the Band component will be seen, e.g., %GRB080916c~\citep{abdo09b}. The actual spectra seen by us are based on the weights of two components. Since the most of GRBs show the non-thermal %components, the non-thermal component frequently wins in this competition.

In our model, in order to obtain the high prompt emission luminosity, we assume that $\gamma_4 \gg \gamma_1$. This assumption is reasonable. This is because estimates based on four methods by Ref.~\cite{rac11} show that the mean observed value of the bulk Lorentz factors of GRB outflows is a few hundred, corresponding to $\gamma_2=\gamma_3\simeq \gamma_1\sim 100$ in our model. Furthermore, within the framework of the collapsar model, a prior relativistic jet-like shell (e.g., shell A) first has to propagate through the envelope of a massive star and clean up almost all of the baryons along the propagation direction of this shell, leaving behind a clean passage for a posterior jet-like shell (e.g., shell B). This, therefore, leads to a reasonable possibility that the Lorentz factor of shell B is much greater than that of shell A.

Usually, we have $\gamma_1 \sim  100$, so $\gamma_4 \sim 10^4 \sim  \gamma ^ 2 _1$ is a universal relationship to obtain the high prompt emission luminosity. Ref.~\cite{yu09} also mentioned that, when $\gamma_4 \sim \gamma ^ 2 _1$, the highest luminosity from internal shocks is expected. In fact, this assumption is not a special case. When collisions among a series of shells with different Lorentz factors occur, the highest luminosity from one collision will cover the others. In other words, we always see the brightest. According to  {Equation~(\ref{eqnum3})}, if deeming that $\gamma_4 \gamma ^ {-2} _1$ does not vary significantly among bursts, we can easily obtain the so-called ``Yonetoku Relation'', $E_p \propto L ^{1/2} _{iso}$~\citep{yone04}, and the ``Amati Relation'', $E_p \propto E ^{1/2} _{iso}$~\citep{amati02}. However, this model is also confronted with some issues, for example, the spectrum is somewhat broad near $E_p$ in contrast to the observed data or the Band function~\cite{yu15}, which can be seen in Figure~\ref{f7}. 

 It is important that we should beware of the empirical Band function. The thermal components and more spectral structures are found in the prompt regimes of some GRBs, which deviate from the so-called Band function~\cite{guiriec11,oganesyan17}. The thermal emission generated by the photosphere is a natural prediction of the generic fireball scenario. The relative strength of thermal emission and non-thermal emission may depend on the various environments~\cite{daigne02,ryde05}. Ref.~\cite{oganesyan17} also claimed that the GRB spectra below the peak energy may present an extra break energy around a few keV, inducing a consistent spectral shape with expectation from the classical synchrotron radiation. More spectral structures of GRBs may make the simple Band function become invalid, and result in an incorrect low-energy spectral index if one forcibly fits them using a Band function. Although our model can present a consistent low-energy spectral index with observations in a certain range, due to the complexities of GRB prompt spectra, more detailed studies are needed.

%Moreover, a relatively large $\gamma_4 \sim 10^4$ is adopted in our numerical calculations, while the posterior shell may move much faster than the prior shell as the circumstellar materials have been swept by the prior shell and the posterior shell moves in a relatively clear environment. These issues need to be addressed through more researches.

%%%%%%%%%%%%%%%%%%%%%%%%%%%%%%%%%%%%%%%%%%
\vspace{6pt} 

%%%%%%%%%%%%%%%%%%%%%%%%%%%%%%%%%%%%%%%%%%
%% optional
%\supplementary{The following are available online at \linksupplementary{s1}, Figure S1: title, Table S1: title, Video S1: title.}

% Only for the journal Methods and Protocols:
% If you wish to submit a video article, please do so with any other supplementary material.
% \supplementary{The following are available at \linksupplementary{s1}, Figure S1: title, Table S1: title, Video S1: title. A supporting video article is available at doi: link.} 

%%%%%%%%%%%%%%%%%%%%%%%%%%%%%%%%%%%%%%%%%%
\authorcontributions{All authors have contributed equally to the preparation of this manuscript.
All authors have read and agreed to the published version of the manuscript.} %MDPI please mentioned every authors name.

\funding{This work was supported by the National Key Research and Development Program of China (grant No. 2017YFA0402600), the National SKA Program of China (grant No. 2020SKA0120300), the National Natural Science Foundation of China under grants No. 11833003, 12003007 and the Fundamental Research Funds for the Central Universities (No. 2020kfyXJJS039).}

\institutionalreview{Not applicable.}

\informedconsent{Not applicable.}

\dataavailability{Not applicable.} 

\acknowledgments{We thank the anonymous referees for the helpful suggestions and comments.}

\conflictsofinterest{The author declares no conflict of interest.} 

\begin{adjustwidth}{-4.6cm}{0cm}

\printendnotes[custom]

\end{adjustwidth}

%%%%%%%%%%%%%%%%%%%%%%%%%%%%%%%%%%%%%%%%%%
\end{paracol}
\reftitle{References}

\end{document}